\documentclass[aps,pra, twocolumn, showpacs,superscriptaddress]{revtex4-1}

\usepackage{graphicx}
\usepackage{SIunits}
\usepackage{pifont}

\begin{document}

\title{A phase-space approach to directional switching in semiconductor ring lasers}

\author{L. Gelens}
\affiliation{Department of Applied Physics and Photonics, Vrije Universiteit Brussel, Pleinlaan 2, B-1050 Brussel, Belgium}
\author{G. \surname{Van der Sande}}
\affiliation{Department of Applied Physics and Photonics, Vrije Universiteit Brussel, Pleinlaan 2, B-1050 Brussel, Belgium}
\author{S. Beri}
\affiliation{Department of Applied Physics and Photonics, Vrije Universiteit Brussel, Pleinlaan 2, B-1050 Brussel, Belgium}
\author{J. Danckaert}
\affiliation{Department of Applied Physics and Photonics, Vrije Universiteit Brussel, Pleinlaan 2, B-1050 Brussel, Belgium}
             
\date{\today}

\begin{abstract} 
We show that a topological investigation of the phase space of a Semiconductor Ring Laser can be used to devise switching schemes which are alternative to optical pulse injection of counter-propagating light. To provide physical insight in these switching mechanisms, a full bifurcation analysis and an investigation of the topology is performed on a two-dimensional asymptotic model. Numerical simulations confirm the topological predictions.
\end{abstract}


\maketitle

\section{Introduction} 
Semiconductor ring lasers (SRLs) have received increasing attention in recent years owing to their possible applications in photonic integrated circuits. This has lead to both theoretical and experimental investigations, ranging from fundamental studies of its nonlinear dynamical behavior to multiple practical applications \cite{SorelOL2002,SorelIEEEJQE2003,LiangAPL1997, AlmeidaOL2004b,HillNature2004, Li_PTL_2008, Wang_PTL_2008, Furst_PTL_2008, Furst_APL_2008, Gelens_OL_2008, Beri_PRL_2008}.
In particular, SRLs attract a lot of interest due to the bistable character of their directional mode operation and the possibility to encode digital information in the direction of emission.  

Recently, all optical flips-flops based on two coupled microrings have been fabricated which could be switched by injection of a counter-propagating signal \cite{HillNature2004}.
On the other hand, a device based on a single ring would benefit of a smaller footprint and could take advantage of the recently demonstrated switching schemes based on one-side injection \cite{Gelens_OL_2008}.
However, single ring devices present a reduced stability, which leads to the appearance of undesired dynamical regimes such as alternate oscillations \cite{SorelOL2002}, or noise-induced mode hopping \cite{Beri_PRL_2008}.

A full understanding of the dynamical stability properties of a SRL as well as an in depth knowledge of the phase space structure would allow for robust devices and smarter switching schemes which rely on the topological properties of the system \cite{Gelens_OL_2008}.
A stability diagram of SRLs has been proposed, which is based on an asymptotic two-dimensional model \cite{VanderSandeJPhysB2008}, including analytical expressions for the local bifurcation curves. However, the important issue of global bifurcations remained unaddressed.

The aim of this paper is twofold. First, we provide a complete bifurcation analysis of the two-dimensional SRL model, which includes all the global bifurcations and the codimension-two points that act as organizing centers for the dynamical behavior.
The phase space topology of the system is systematically investigated for time scales which are slower than the relaxation oscillations \cite{VanderSandeJPhysB2008}. In particular, we identify the role of the different physical parameters on the global shape of the invariant manifolds.
Secondly, we propose two novel topology-based schemes to switch the operation of the ring from clockwise (CW) to counter-clockwise (CCW) mode. Such schemes are based on externally steering the system in the phase space and are alternative to the standard approach of injecting an optical signal counter-propagating to the lasing mode \cite{YuanIEEE2008, PerezOE2007}. More specific, we will show that it is possible to switch in a robust way by current and phase modulation.

The paper is structured as follows.
In Sect.\ \ref{Sect::RE_general}, we start by reviewing the existing models that has been shown to adequately describe the dynamical behavior of SRLs \cite{SorelOL2002,VanderSandeJPhysB2008} and we revisit previous results on injection induced switches in SRLs \cite{YuanIEEE2008, PerezOE2007, Gelens_OL_2008}. 
In Section \ref{Sect::Bifurcations}, we provide insight in the full bifurcation picture of a two-dimensional SRL model, including the global bifurcation lines which were not addressed before.
In Section \ref{Sect::Manifolds}, we study the shape of the invariant manifolds of the saddle points in reduced SRL system and we discuss the effect of the individual parameters on the topology of the system. 
The particular shape of these invariant manifolds allows us to devise novel switching schemes through current modulation and phase modulation, which we discuss in Sect. \ref{Sec::Switching}. Our predictions are verified numerically based on the full rate equation model. Finally, in Sect. \ref{Sec::Discussion} we draw conclusions of our analysis and discuss possible future work.

\section{Semiconductor ring laser models}\label{Sect::RE_general}
We consider a SRL operating in a single-longitudinal, single-transverse mode. In order to describe the dynamical behavior of this SRL, we will focus on a rate-equation approach allowing for an easy and fast numerical implementation. Problems involving wavelength changes, however, fall outside the scope of such rate-equation models, and a traveling wave model would be more suitable to tackle such issues \cite{Javaloyes_JQE_2008, Radziunas_SPIE_2008}.\\
In the limit of small outcoupling from the ring cavity, the total electric field oscillating in the ring can be written as the sum of two counter-propagating waves:
\begin{eqnarray}
E(z,t) = E_1(t) \exp\left[i\left(\omega_0 t - k_0 z\right)\right] \nonumber \\
+ E_2(t) \exp\left[i\left(\omega_0 t + k_0 z\right)\right] + \text{c.c.}.
\end{eqnarray}
Here $k_0$ is the longitudinal wavenumber and $\omega_0$ is the optical frequency of the mode. In the slowly varying envelope approximation, the amplitudes $E_{1,2}$ vary on time scales which are slower than $2 \pi / \omega_0$.
The rate-equation model is formulated mathematically in terms of two rate equations for the slowly varying amplitudes $E_{1,2}$ and one rate equation for the carrier number $N$. The equations read \cite{SorelOL2002,SorelIEEEJQE2003,VanderSandeJPhysB2008}: 
\begin{eqnarray}
\dot{E}_{1,2} &= \kappa(1+i\alpha)\left[N\left(1-s|E_{1,2}|^2-c|E_{2,1}|^2\right)-1\right]E_{1,2} \nonumber \\
&- k e^{i \phi_k}E_{2,1}, \label{Eq::Field1::Original}\\
\dot{N} &= \gamma [ \mu -N - N\left(1-s|E_1|^2-c|E_2|^2\right)|E_1|^2 \nonumber \\
&- N\left(1-s|E_2|^2-c|E_1|^2\right)|E_2|^2] , \label{Eq::Carriers::Original}
\end{eqnarray}
where dot represents differentiation with respect to time $t$, $\kappa$ is the field decay rate, $\gamma$ is the carrier decay rate, $\alpha$ is the linewidth enhancement factor and $\mu$ is the renormalized injection current with $\mu\approx0$ at transparency and $\mu\approx1$ at lasing threshold. The two counter-propagating modes are considered to saturate both their own and each other gain due to e.g. spectral hole burning effects. Self- and cross-saturation effects are added phenomenologically and are modeled by $s$ and $c$. For a realistic device cross saturation is stronger than self-saturation.

Contrary to the case of solid-state lasers \cite{Zeghlache83}, in the devices under study, the standing-wave pattern has a spatial period much smaller than the carrier diffusion length. Therefore, longitudinal variations of the carrier density are washed out by the diffusion. As a result, the dynamics of such a carrier grating can be neglected \cite{SorelIEEEJQE2003, Etrich92, Khandokhin95}. However, a linear coupling between the two counter-propagating modes remains. Reflection of the counter-propagating modes occurs at the point where light is coupled out of the ring cavity into a coupling waveguide and can also occur at the end facets of the coupling waveguide. These localized
reflections result in a linear coupling between the two fields characterized by an amplitude $k$ and a phase shift $\phi_k$ \cite{Spreeuw90}. 

In order to gain a deeper insight in the switching mechanism, we have proposed an asymptotic simplification of Eqs.\ (\ref{Eq::Field1::Original})-(\ref{Eq::Carriers::Original}) \cite{VanderSandeJPhysB2008}. A similar reduction has been proposed to describe the polarization switching behavior in VCSELs \cite{ErneuxPRA1999}. On timescales slower than the relaxation oscillations, it can be shown that the total intensity is conserved:
\begin{eqnarray}
|E_1|^2+|E_2|^2 = \mu-1 > 0. \label{Eq::ConservationLaw}
\end{eqnarray}
The slow timescale dynamics is then described by the time evolution of two auxiliary angular variables:
\begin{eqnarray}
\theta' &= -2\sin\phi_k\sin\psi + 2 \cos\phi_k\cos\psi\sin\theta \nonumber \\
& + J \sin\theta\cos\theta , \label{Eq::Theta}\\
\cos\theta \psi' &= \alpha J \sin\theta\cos\theta + 2 \cos\phi_k \sin\psi \nonumber \\
&+ 2 \sin\phi_k \cos\psi \sin\theta. \label{Eq::Psi}
\end{eqnarray}
where $\theta = 2 \arctan \sqrt{|E_2|^2 / |E_1|^2} - \pi/2 \in [-\pi/2,\pi/2]$ represents the relative modal intensity and $\psi \in [0,2\pi]$ is the phase difference between the counter-propagating modes. Prime now denotes derivation to the slow timescale $\tau=k t$.  Finally, in this reduced model the pump current has been rescaled as 
\begin{equation}
J=\kappa (c-s)(\mu-1)/k .\label{Eq::CurrentJ}
\end{equation}
As the phase space of Eqs.~(\ref{Eq::Theta})-(\ref{Eq::Psi}) is restricted to two dimensions, it allows for a clear physical picture of the influence of all parameters on the dynamical evolution of the variables in a plane. Remark that further simplifications to a one dimensional model, as was possible in some cases for VCSELs \cite{Willemsen00a, VanderSande_PRA_2003}, are not valid for SRLs \cite{Gelens_OL_2008, Beri_PRL_2008}.


Previously \cite{Gelens_OL_2008}, we have reported on the possibility of directional mode switching in semiconductor ring lasers through optical injection co-propagating with the lasing mode. This mechanism is counter intuitive as the mode receiving the least energy from the optical injection prevails after the injection pulse, but can be explained due to the folded shape of the stable manifold of a saddle point $S$ that is present in the system.\\

\begin{figure}[t!]
\centering
\includegraphics[width=\columnwidth]{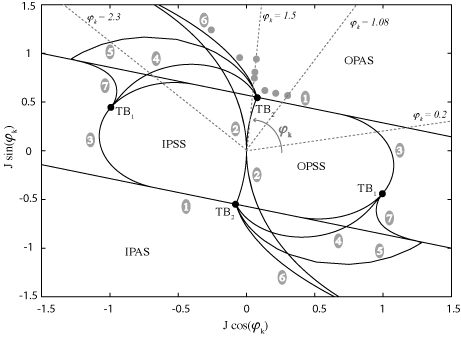}
\caption{\label{fig:StabilityDiagram} Stability diagram for a fixed value $\alpha=3.5$ showing the different stable operation regimes for $J>0$. The different black solid lines indicate the different bifurcation currents, and are depicted in a polar plot $(J, \phi_k)$. The dotted lines depict four cuts for four different values of $\phi_k$, for which we show the corresponding PI-curves in Fig.\ \ref{Fig::BifurcationPosJ}. The eight solid grey dots denote the different operating points of which we plot the invariant manifolds in Figs.\ \ref{Fig::sMan_J}, \ref{Fig::sMan_Phi}, \ref{Fig::sMan_Alpha}. Finally, the denotations 1-6 correspond to (1) the Pitchfork bifurcation of the OPSS (IPSS), (2) Hopf bifurcation of the OPSS (IPSS), (3) Saddle-Node bifurcation the OPSS (IPSS), (4) Sub-critical Hopf bifurcation the OPAS (IPAS), (5) Glueing bifurcation, (6) a Fold of Cycles, and (7) a double homoclinic bifurcation. Multiple Takens-Bogdanov (TB) co-dimension two points are also depicted.}
\end{figure}

The understanding of this novel feature in ring lasers is thus largely based on the specific structure of the two-dimensional asymptotic phase space, in particular the shape of the invariant manifolds of the saddle point. In this manuscript, we wish to extend this principle by introducing new ways to switch the SRL between both its directional modes. Exploiting specific phase-space structures, we will demonstrate the possibility of directional switching through current and phase modulation. In order to be able to adequately design these schemes, it is necessary to have insight into how the phase space picture changes with the relevant parameters. Therefore, in what follows, we will first perform a full bifurcation analysis of the reduced model, which afterwards will provide us with the optimal parameter regions for an in-depth study of these different switching schemes.

\section{Bifurcation analysis} \label{Sect::Bifurcations}

One advantage of the asymptotic reduction presented in Ref.\ \cite{VanderSandeJPhysB2008} is the possibility to have an analytical expression for all the steady state solutions of the system. 
The second advantage is the two-dimensionality of its phase space, which enables not only an analytical study of all local bifurcations, but also simplifies the numerical continuation techniques necessary for obtaining global bifurcations. Furthermore, it allows for an easy visualization of the dynamical behavior in phase-space.

Two different types of stationary solutions exist. On the one hand, solutions corresponding to bi-directional emission ($\theta=0$) exist for all model parameters and are identified by their value of $\psi$. For $\psi=0$ and $\psi=\pi$, we can distinguish between the in-phase symmetric-solution (IPSS) and the out-of-phase symmetric solution (OPSS). On the other hand, asymmetric solutions corresponding to unidirectional emission can emerge from either a pitchfork bifurcation of the symmetric solutions or from a saddle-node bifurcation and are given by:
\begin{eqnarray}
\theta(\psi) &=& \arcsin\left( \frac{\alpha \sin \phi_k + \cos \phi_k}{\alpha \cos \phi_k - \sin \phi_k}  \tan \psi \right), \\
J\left( \psi \right) &=& 2 \csc \theta(\psi) \sec \theta(\psi)  \\
 && \left[\sin \phi_k \sin \psi - \cos  \phi_k \cos \psi \sin \theta(\psi) \right]. \nonumber 
\end{eqnarray}
If the asymmetric solutions originates from a pitchfork from the IPSS (OPSS), we will refer to them as IPAS (OPAS). This pitchfork bifurcation current can be easily obtained in an analytical form:
\begin{eqnarray}
J^{PF} =  \frac{\pm 2}{\cos \phi_k + \alpha \sin \phi_k}, \label{Eq::PFBifurcation}
\end{eqnarray} 
where the upper (lower) sign again corresponds to the IPSS (OPSS). We will study the stability of the system in a two parameter plane defined by $J$ and $\phi_k$ with a fixed $\alpha$. If we choose a polar representation where $x=J \cos \phi_k$ and $y=J \sin \phi_k$, the pitchfork bifurcation denoted by \ding{202} in Figure \ref{fig:StabilityDiagram} is a straight line $x + \alpha y = \pm 2$. For a certain parameter regime, the IPAS (OPAS) are born in a saddle node bifurcation or fold bifurcation before the pitchfork bifurcation has occurred. This Saddle-Node bifurcation is given analytically by:
\begin{eqnarray}
\cos^2 \psi^{SN} &=& \pm 2 \frac{ \cos \phi_k + \alpha \sin \phi_k}{1+\alpha^2} \label{Eq::FoldBif1}
\end{eqnarray}
and it is denoted by \ding{204} in Figure \ref{fig:StabilityDiagram}.

The symmetric solutions can also change stability through a Hopf bifurcation at $J= J_{H}(\phi_k) $ of the symmetric solutions (OPSS, IPSS) as described by the expression below:
\begin{eqnarray}
J^{H} = \pm 4 \cos \phi_k , \label{Eq::HopfBifurcation}
\end{eqnarray} 
where the upper (lower) sign corresponds to the IPSS (OPSS). In Fig.\ \ref{fig:StabilityDiagram}, these Hopf bifurcations lines are segments of  the circles $[(x\pm2)^2 + y^2 = 4]$ denoted by \ding{203}. These Hopf bifurcations lines end at the pitchfork bifurcations in a Takens-Bogdanov point with symmetry (TB$_2$) \cite{Guckenheimerbook}.

The asymmetric solutions can change stability through a subcritical Hopf bifurcation given by
\begin{eqnarray}
\tan^2 \psi^H_{AS} &=& \left(\frac{\sin \phi_k - \cos \phi_k\alpha }{\cos \phi_k + \sin \phi_k\alpha } \right)^2 \times\\
 && \frac{\sin^2 \phi_k - \cos^2 \phi_k -   \alpha \sin 2\phi_k}{ 3\sin^2 \phi_k + \cos^2 \phi_k -   \alpha \sin 2\phi_k}. \nonumber
\end{eqnarray}
These Hopf bifurcations (\ding{205} in Fig.\ \ref{fig:StabilityDiagram}) emerge from a Takens-Bogdanov point (TB$_1$) at the Saddle-Node bifurcation line.

\begin{figure}[]
\centering
\includegraphics[width=\columnwidth]{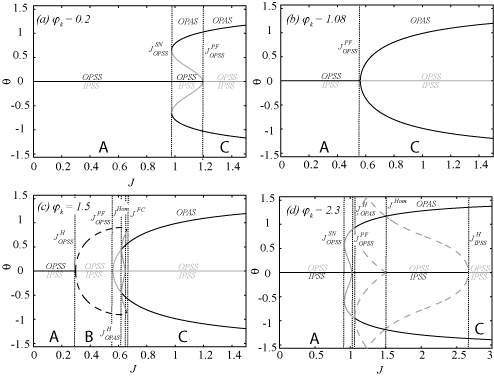}
\caption{\label{Fig::BifurcationPosJ} Bifurcation diagrams of Eqs.\ (\ref{Eq::Theta})-(\ref{Eq::Psi}) depicting the extremes of $\theta$ vs.\ injection current $J$. The steady state values of $\theta$ are denoted by full lines, while the extrema of periodically oscillating $\theta$ are indicated with dashed lines. Black (grey) color is used for stable (unstable) fixed points or limit cycles. The dotted lines represent the different bifurcation currents, and are denoted as explained in the text. Black (grey) OPSS/IPSS/OPAS denotations correspond to stable (unstable) OPSS/IPSS/OPAS. From (a) to (d), we have $\phi_k =$ $0.2$, $1.08$, $1.5$ and $2.3$, respectively. $\alpha=3.5$.}
\end{figure}

As previously mentioned, the dynamics is organized by two co-dimension two Takens-Bogdanov points. It is known that from these points there should also be global bifurcations emerging \cite{Kuznetsovbook}. Therefore, we have complemented our analytical analysis of the bifurcations in the reduced system, with a numerical approach using the continuation software package AUTO \cite{AUTO97}. From TB$_{1}$, the point where the Saddle-Node bifurcation line \ding{204} and the Sub-critical Hopf line \ding{205} meet, emerges a double homoclinic bifurcation line \ding{208}. From TB$_2$, the point where the Super-critical Hopf bifurcation line \ding{203} and the Pitchfork bifurcation \ding{202} meet, emerges a double homoclinic connection with a glueing bifurcation \ding{206} and a fold of cycles \ding{207}. Finally, also the Super-critical Hopf bifurcation and the Fold of Cycles coincide in a co-dimension 2 point called a Bautin bifurcation point (not shown) \cite{Kuznetsovbook, Guckenheimerbook}.

As the Takens-Bogdanov points involve local bifurcations, we have obtained analytical expressions. The critical phase $\phi_k^{TB_2}$ of this Takens-Bogdanov ($TB_2$) point with symmetry can be expressed analytically as follows: 
\begin{eqnarray}
\sin^2(\phi_k^{TB_2}) &=&  \frac{1}{2}  \pm \frac{\alpha }{2 \sqrt{\alpha^2 + 1}}, \\
J^{TB_2} &=& \pm 4 \cos \phi_k^{TB_2}. 
\label{Eq::TB_2}
\end{eqnarray}

We point out that Figure \ref{fig:StabilityDiagram} provides a complete picture of the different dynamical regimes for $\alpha = 3.5$, but only for positive values of $J$, meaning that $c>s$, being the physically relevant parameter region.  The bifurcation lines for negative values of $J$, however, are exactly the same, but due to the particular invariance below of the reduced system (Eqs.\ \ref{Eq::Theta}-\ref{Eq::Psi}) the stability is reversed: 
\begin{eqnarray}
\phi_k &&\rightarrow \phi_k + \pi, \\
J  &&\rightarrow -J, \\
t &&\rightarrow -t.
\label{Eq::invariant_tfo}
\end{eqnarray}

While performing a bifurcation analysis, both analytically and numerically, many different dynamical regions of operations can be found in the reduced system Eqs.~(\ref{Eq::Theta})-(\ref{Eq::Psi}), as can be seen from the different PI-curves in Fig.\ \ref{Fig::BifurcationPosJ}. Three different regions of operation A,B,C can be identified in Fig.~\ref{Fig::BifurcationPosJ}(a-d) according to the relative magnitude of the linear and nonlinear coupling terms between CW and CCW mode.
\\
At the threshold current, $J\approx0$, laser action starts. 
When operating close to the threshold, at low power, 
the nonlinear coupling between CW and CCW is negligible when compared to the linear coupling induced by the backscattering. The resulting operation is bidirectional with CW and CCW operating either in-phase or out-of-phase according to $\phi_k$.
(see region A in Figure \ref{Fig::BifurcationPosJ}). 
For large values of the pump parameter $J$, the nonlinear gain saturation becomes the dominant coupling mechanism between the modes and two out-of-phase (in-phase) asymmetric OPAS (IPAS) solutions become possible.
The optical output power is mainly concentrated in one propagation direction, called unidirectional operation. Because of the device symmetry two OPAS exist: one where $|E_1|^2>|E_2|^2$ and vice versa. In this regime, the device exhibits bistability (See Regime C in Figure \ref{Fig::BifurcationPosJ}).
For intermediate values of $J$, the linear and nonlinear coupling are comparable (Region B) and the dynamics of the system as well as the shape of the P-I curve depend on the value of $\phi_k$. A SRL aimed to application such as all-optical information storage is expected to be operated at the edge between region B and C. 
\\
Depending on the linear coupling phase $\phi_k$, the dynamical behavior that the SRL exhibits at the transition between bidirectional and unidirectional operation, originates from instabilities of either the out-of-phase or in-phase solutions. The stable bidirectional solution can be destabilized, immediately going to stable unidirectional operation [see Fig.\ \ref{Fig::BifurcationPosJ}(b)], but there can be many other possibilities of which we show a couple of examples in the other panels of Fig.\ \ref{Fig::BifurcationPosJ}. E.g.\ in Fig.\ \ref{Fig::BifurcationPosJ}(a), there exists a region where both the stable bidirectional solution as the stable unidirectional solution coexist. In Fig.\ \ref{Fig::BifurcationPosJ}(c), the destabilisation of the bidirectional solution leads to region B, characterized by stable alternate intensity oscillations between the two counter-propagating modes, which has been observed experimentally \cite{SorelOL2002}. At this Hopf bifurcation point, a limit cycle representing a dynamic competition between the two counter-propagating modes appears. Furthermore, a combination of the above situations is also possible. For instance, in Fig.\ \ref{Fig::BifurcationPosJ}(d), there exists tristability between the stable IPSS and the stable OPAS, with the presence of different unstable limit cycles separating the different basins of attraction of these stable solutions. 

\section{Phase-space representation}\label{Sect::Manifolds}

In this section we investigate the phase space $\left( \theta, \psi \right)$ for a SRL operating in the bistable regime. In this bistable unidirectional regime, Eqs.\ (\ref{Eq::Theta})-(\ref{Eq::Psi}) have four stationary solutions: an unstable in-phase (out-of-phase) bidirectional state in $(0,0)$ [$(0,\pi)$]; two symmetric stable states {\bf $CW$} (clockwise) and {\bf $CCW$} (counter-clockwise) at $\psi \approx \pi$ ($\psi \approx 0$)  both corresponding to undirectional operation, and a saddle point {\bf S} in $(0 , \pi)$ [$(0 , 0)$] which is the unstable out-of-phase (in-phase) bidirectional solution. As we are interested in operating regimes which are relevant for optical information storage, we consider here parameter values such that we are at the edge between region B and C (see Fig.\ \ref{Fig::BifurcationPosJ}). In particular, we focus on parameters such that we are relatively close to the Fold of Cycles bifurcation line \ding{207}. 

We are interested in the qualitative shape of the invariant manifolds of the saddle-point {\bf S} and their dependence on the principal parameters of the system. In particular, we focus on the stable manifold of {\bf S} separating the basins of attractions of the {\bf $CW$} and {\bf $CCW$} mode, shown in black and white in Figs.\ \ref{Fig::sMan_J}, \ref{Fig::sMan_Phi}, \ref{Fig::sMan_Alpha} (The parameter sets under study in these figures are depicted by grey dots in Fig.\ \ref{fig:StabilityDiagram}).

\subsubsection{Influence of $J$}
\begin{figure}[]
\centering
\includegraphics[width=\columnwidth]{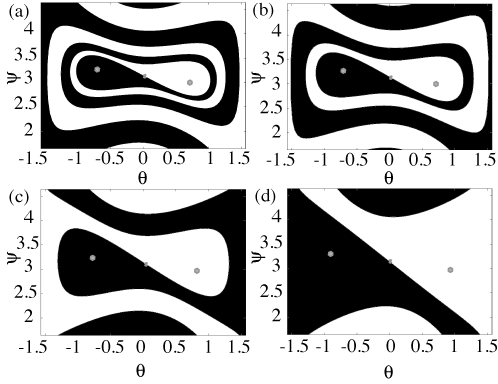}
\caption{\label{Fig::sMan_J} Stable manifold at $\phi_k = 1.5$, and $\alpha=3.5$ for three increasing different values of the normalized current $J$: $J=0.71$, $0.739$, $0.8$, $0.93$, from (a) to (d). The basins of attractions of the {\bf $CW$} and {\bf $CCW$} mode are given by the black and white region.}
\end{figure}

We study first the dependence of the phase space structures on the pump parameter $J$ [see Eq.\ (\ref{Eq::CurrentJ})]. As defined in Sec.~\ref{Sect::RE_general}, $J$ measures the ratio of nonlinear and linear coupling between {\bf $CW$} and {\bf $CCW$} modes.
Consider here ($\phi_k$, $\alpha$) = ($1.5, 3.5$); when varying the parameter $J$, one observes three different regions of operation: bidirectional operation (A), bidirectional operation with alternating oscillations (B) and unidirectional operation (C) 
[see Fig.~\ref{Fig::BifurcationPosJ}] as it has been experimentally observed \cite{SorelOL2002}. 
The shape of the stable manifold of {\bf S},  and the basins of attraction of {\bf $CW$} and {\bf $CCW$} are depicted in Fig. \ref{Fig::sMan_J}. 
The basins of attraction of $CW$ and $CCW$ fold into each-other as the stable manifold of $S$ spirals inwards. As $J$ increases, the stable manifold unfolds and the basins of attraction separate. Therefore, a large bias current $\mu$ or a small coupling amplitude $k$ correspond to unfolded basins of attraction.

\subsubsection{Influence of $\phi_k$}

\begin{figure}[]
\centering
\includegraphics[width=\columnwidth]{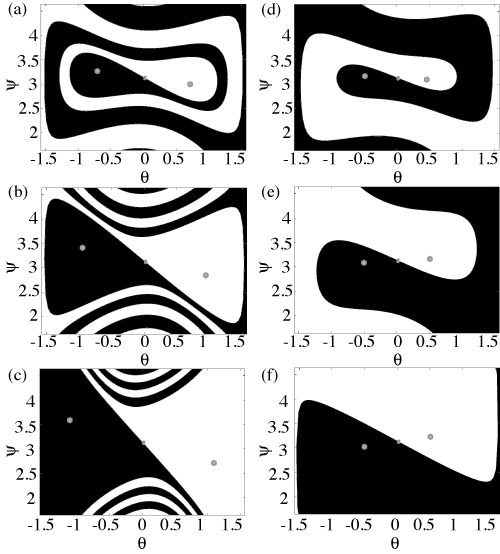}
\caption{\label{Fig::sMan_Phi} Stable manifold at $\alpha=3.5$ for different values of the linear coupling phase $\phi_k$: from (a) to (f) ($\phi_k$, $J$) $=$ $(1.5, 0.739)$, $(1.64, 0.972)$, $(1.78, 1.260)$, $(1.36, 0.625)$, $(1.22, 0.625)$, $(1.08, 0.636)$. The basins of attractions of the {\bf $CW$} and {\bf $CCW$} mode are given by the black and white region.}
\end{figure}

It is very hard to determine the value of $\phi_k$ for a SRL device as this parameter describes the phase of the backscattering that arises due to different phenomena: localized backscattering at the output coupler region, at the end facets of the output waveguides and distributed backscattering originating from e.g. side-wall surface roughness in the ring itself. Therefore, it is useful to gain insight in the effect of the linear coupling phase $\phi_k$ in the entire region from zero to $2\pi$. In order to study the influence of $\phi_k$ on the shape of the manifold, we will focus on the region having the stable OPAS at higher current ($\phi_k \in [0,\pi]$). The region $\phi_k \in [\pi, 2\pi]$ is exactly the same if one interchanges everywhere in-phase (IP) and out-of-phase (OP). \\
One could wonder whether this folded shape of the stable manifold is only present if one has a region of alternating oscillations in the PI scan, as was the case in Fig.\ \ref{Fig::sMan_J}. We have checked the evolution of the manifold shape in the region $\phi_k \in [0, \phi_k^{TB_2}]$ where one does not have a Hopf bifurcation [See Fig.\ \ref{Fig::sMan_Phi}(d)-(f)]. $\phi_k^{TB_2}$ defines the phase $\phi_k$ where the second Takens-Bogdanov bifurcation point lies.\\
For $\phi_k \in [0, \phi_k^{TB_2}]$, one can notice in Fig.\ \ref{Fig::sMan_Phi}(d)-(f) that although one does not observe oscillations in a PI scan, the manifold can still have a folded shape. It unfolds continuously with decreasing values of $\phi_k$.\\
In the region $\phi_k  > \phi_k^{TB_2}$, many different bifurcation scenarios can take place in a PI scan. However, the bifurcation scheme always ends in a Fold of Cycles where a stable limit cycle (originating from either the OPSS or the IPSS) touches the unstable limit cycle formed in the glueing bifurcation. As becomes clear from Fig.\ \ref{Fig::sMan_Phi}(a)-(c), the manifold unfolds itself with increasing values of $\phi_k$. Due to the fact that the distance between the glueing bifurcation and the fold of cycles also increases, the stable manifold is already more unfolded by the time the fold of cycles occurs.

\subsubsection{Influence of $\alpha$}

\begin{figure}[]
\centering
\includegraphics[width=\columnwidth]{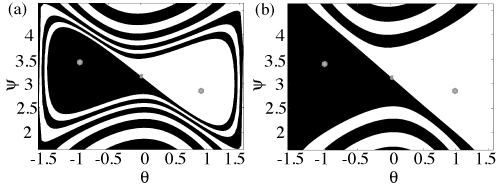}
\caption{\label{Fig::sMan_Alpha} Stable manifold at ($\phi_k$, $J$) $=$ $(1.64, 0.972)$ for different values of the linewidth enhancement factor $\alpha$: (a) $3.3$, (b) $3.7$. The basins of attractions of the {\bf $CW$} and {\bf $CCW$} mode are given by the black and white region.}
\end{figure}

Also the value of the line-width enhancement factor $\alpha$ is difficult to determine for a practical device \cite{Zanola_PTL_2008}. How $\alpha$ affects the dynamical system [Eqs.~(\ref{Eq::Theta})-(\ref{Eq::Psi})] is illustrated in Fig.\ \ref{Fig::sMan_Alpha}. We have plotted the manifold shapes for different values of the linewidth enhancement factor $\alpha$, while keeping ($\phi_k$, $J$)=$(1.64, 0.972)$. We find that increasing values of $\alpha$ lead to an unfolding of the stable manifold for $\phi_k  > \phi_k^{TB_2}$, while for $\phi_k \in [0, \phi_k^{TB_2}]$, the reverse behavior is observed.

This can be explained with the analytical expression for the Pitchfork bifurcation of Eq.\ (\ref{Eq::PFBifurcation}). This Pitchfork bifurcation is a straight line in Fig.\ \ref{fig:StabilityDiagram} with a slope given by $-1/\alpha$. Thus, on the one hand increasing the value of $\alpha$ for $\phi_k  > \phi_k^{TB_2}$ will shift all the bifurcation lines further away from the operating point ($\phi_k$, $J$) $=$ $(1.64, 0.972)$, such that the manifold unfolds itself. On the other hand, increasing the value of $\alpha$ for $\phi_k \in [0, \phi_k^{TB_2}]$, the pitchfork bifurcation will come closer to the chosen ($\phi_k$, $J$) point in the bistable regime, such the manifold will spiral more into itself.

In conclusion, in this Section we have revealed a non-trivial shape of the basin of attraction of the system which is not observed in other optical systems such as VCSELs \cite{Willemsen00a}. 
The standard approach to switches, consisting in injection of counter-propagating signal \cite{YuanIEEE2008, PerezOE2007} correspond, in phase-space, to steering the system very close to the final counter-rotating state. However, since Figs.~\ref{Fig::sMan_J}-\ref{Fig::sMan_Phi} suggest that other regions of the basin of attraction are easier or more useful to target, we investigate this possibility in more depth in the next Section.

\section{Alternative switching mechanisms}\label{Sec::Switching}
In this section, we propose different switching mechanisms which are based on the structure of the phase space of the system. Consider for instance a SRL operating in the CCW mode. In order to switch operating direction, one needs to steer the system to the basin of attraction of the CW state with an external perturbation such as a pulse in the bias current or an external optical field.
One can engineer switches in the following way. Consider the laser operating stationary in unidirectional regime, for instance CW $\left( \theta_{CW}, \psi_{CW} \right)$, and introduce a perturbation in the system which shifts the system away from this CW state $\left(\theta_{CW}, \psi_{CW} \right)$. It is clear from the previous discussion that it is possible to use such perturbation to drive the system into the basin of attraction of the opposite state. Once the perturbation is removed, the system will deterministically relax to CCW operation. The switching induced by the injection of a resonant optical signal has been investigated in \cite{YuanIEEE2008, PerezOE2007, Gelens_OL_2008} and we will not discuss it further here.
We propose to induce a switch by:
\begin{enumerate}
\item modulation of the pump current $\mu$,
\item external modulation the linear coupling phase $\phi_k$.
\end{enumerate}

\subsection{Current modulation}\label{Sec::CurrentMod}

\begin{figure}[]
\centering
\includegraphics[width=\columnwidth]{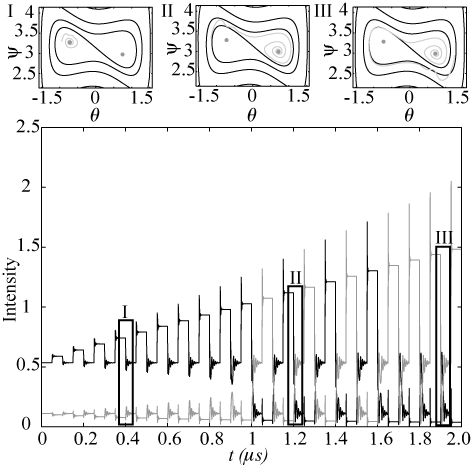}
\caption{\label{Fig::CurrentMod} Numerical simulations of Eqs.\ (\ref{Eq::Field1::Original})-(\ref{Eq::Carriers::Original}), where the current is modulated around $\mu_{bias} = 1.65$ ($J_{bias} = 0.739$). The current modulation $\Delta \mu$ is continuously changed in time by $0.044 m$ with a period of $T = 100 ns$, where $m$ is an integer describing the $m^{th}$ modulation step. The insets $I, II, III$ depict the evolution of the system projected in the two-dimensional phase plane ($\theta$, $\psi$) for three values of the current modulation, respectively for $m = 4, 12$ and $19$. $\phi=1.5$, $k=0.44 ns^{-1}$, $\alpha=3.5$, $c=0.01$, $s=0.005$, $\kappa = 100 ns^{-1}$, $\gamma = 0.2 ns^{-1}$.}
\end{figure}

Knowing that the OPAS separate for increasing values of the normalized pump current $J$, one possibility to induce a switch is by increasing the pump current for a certain time, such that the system relaxes to the new OPAS state with higher intensity. If one then decreases the current fast enough, the system will find itself in the basin of attraction of one of the original lower intensity OPAS modes. It will follow the deterministic trajectories of this last current value and relax to the targeted OPAS state. Which one depends on the amplitude of the current modulation determining where in the stable manifold one ends up. 

A confirmation of this behavior can be seen in Fig.\ \ref{Fig::CurrentMod}, where the full SRL rate equation model [Eqs.\ (\ref{Eq::Field1::Original})-(\ref{Eq::Carriers::Original})] is simulated for $\mu_{bias} = 1.65$ ($J = 0.739$), $\phi=1.5$, $k=0.0044$, $\alpha=3.5$, $c=0.01$, $s=0.005$, $\kappa = 100 ns^{-1}$, $\gamma = 0.2 ns^{-1}$. Furthermore, a time-varying modulation $\Delta \mu(m)$ is applied with a period $T = 100 ns$. $\Delta \mu(m)$ is given by $0.044 m$, where $m$ is an integer describing the $m^{th}$ modulation step. This $\Delta \mu$ corresponds to $\Delta J = 0.05 m$. In Fig.\ \ref{Fig::CurrentMod}, one can clearly see that there exist regions of $\Delta \mu$ inducing a switch, and others that do not, confirming the relevance of the two-dimensional stable manifold.\\

The insets $I, II, III$ in Fig.\ (\ref{Fig::CurrentMod}) depict the evolution of the system projected in the two-dimensional phase plane ($\theta$, $\psi$) for three values of the current modulation, respectively for $m = 4, 12$ and $19$. These insets demonstrate that depending on your initial condition (determined by $\Delta \mu$), you either switch to the opposite mode or not. More importantly, one can see that the invariant manifolds of the saddle point as obtained from the reduced 2D model are followed by the system described with the full SRL rate equation model, confirming the validity of the asymptotically reduced system. However, one should note that this good correspondence between the full rate-equation system and the reduced one is only obtained after a certain transient time, determined by the damping rate of the relaxation oscillations (ROs). We have indeed already mentioned before that the two-dimensional system [Eqs.~(\ref{Eq::Theta})-(\ref{Eq::Psi})] is only valid on a slow time scale ($\tau = k t$), such that the faster ROs should have first damped out before one can adequately predict the evolution of the SRL with the reduced model.\\
In Fig.\ \ref{Fig::CurrentMod}, we have taken an infinitely fast rise and fall time for the current modulation, which is not realistic from an experimental point of view. 
In order to understand the effect of a finite rise/fall time, we have investigated the case of $\Delta \mu = 0.0528$ (Inset II in Fig.\ \ref{Fig::CurrentMod}) for different values of the fall time. 
Our simulations (not shown) reveal that fall times which are longer that a critical $t_{fall} \approx 1.35 ns$ do not result in a switch, and the system returns to the original state by following the instantaneous equilibrium states.

\subsection{Phase modulation}

\begin{figure}[h!]
\centering
\includegraphics[width=\columnwidth]{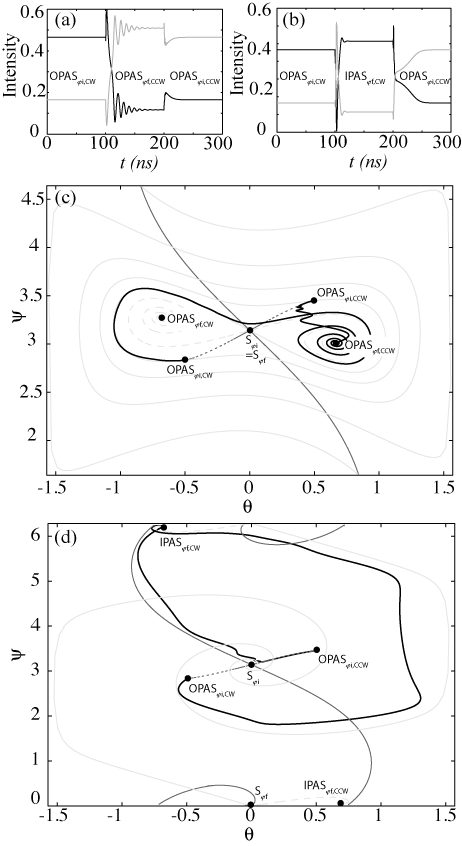}
\caption{\label{Fig::PhaseMod} Numerical simulations of Eqs.\ (\ref{Eq::Field1::Original})-(\ref{Eq::Carriers::Original}), where the linear coupling phase is modulated between $\phi_{k, i} = 0.7$ and (a,c) $\phi_{k, f} = 1.5$, (b,d) $\phi_{k, f} = 4.3$ with a period $T = 200 ns$. (a,b) depict a timetrace demonstrating the possibility to switch the SRL between the two OPAS. (c,d) show the projection of (a,b) in the phase portrait of the system in a solid black line. $CW$ and $CCW$ are the stable unidirectional states of the system; $S$ is the saddle. The stable (unstable) manifold of $S$ is indicated with a grey solid (dashed) line for the two values of the phase $\phi_{k, i}$ and $\phi_{k, f}$, respectively in dark grey and light grey.  $\mu= 1.63$ ($J = 0.716 $), $\phi=1.5$, $k=0.44 ns^{-1}$, $\alpha=3.5$, $c=0.01$, $s=0.005$, $\kappa = 100 ns^{-1}$, $\gamma = 0.2 ns^{-1}$. }
\end{figure}

An important drawback of using current modulation to induce directional switches in a SRL, is that it should be feasible to electrically modulate the pump current with fall times of the order of $ns$. Modulating the linear coupling phase could help to circumvent this issue, since the modulation could be done optically. Moreover, it is clear from Fig.~\ref{fig:StabilityDiagram}, that one can induce larger effects by modulating the phase as compared with current modulation and therefore it is more flexible. In what follows, we propose two switching schemes that are both based on steering induced by phase modulation. Consider for instance injection of a non resonant external optical signal in the coupler. Far from the resonance condition, the main effect of such signal would be to alter the refractive index of the coupler without altering the field inside the ring cavity. In first approximation, for a standard semiconductor, we can disregard the changes in $k$ and consider only changes in $\phi_k$.

In Fig.\ \ref{Fig::PhaseMod}, we show numerical simulations of the full rate-equation system [Eqs.\ (\ref{Eq::Field1::Original})-(\ref{Eq::Carriers::Original})], where the linear coupling phase is modulated between $\phi_{k, i} = 0.7$ and $\phi_{k, f} = 1.5$ [Fig.\ \ref{Fig::PhaseMod}(a),(c)], $\phi_{k, f} = 4.3$ [Fig.\ \ref{Fig::PhaseMod}(b),(d)] with a period $T = 200 ns$. Fig.\ \ref{Fig::PhaseMod}(a)-(b) depict a timetrace demonstrating the possibility to switch the SRL between the two OPAS. Fig.\ \ref{Fig::PhaseMod}(c)-(d) show the projection of Fig.\ \ref{Fig::PhaseMod}(a)-(b) in the two-dimensional phase space of the reduced system in a solid black line. $CW$ and $CCW$ denote the stable unidirectional states of the system; $S$ is the saddle. The stable (unstable) manifold of $S$ is indicated with a grey solid (dashed) line for the two values of the phase $\phi_{k, i}$ and $\phi_{k, f}$, respectively in dark grey and light grey. The other parameters are kept fixed: $\mu= 1.63$ ($J = 0.716 $), $\phi=1.5$, $k=0.0044$, $\alpha=3.5$,$c=0.01$, $s=0.005$, $\kappa = 100 ns^{-1}$, $\gamma = 0.2 ns^{-1}$.\\
One can distinguish between two possible methods to switch the laser between two OPAS (IPAS) states through phase modulation. The first one is based on changes in $\phi_k$ smaller than $\pi$ $(\Delta \phi_k < \pi)$ and do not largely affect the variable $\psi$. Therefore, they involve only the areas of phase space which are close to the initial modal phase difference $\psi = \pi$ ($\psi = 0$). 
The second method takes advantage of a phase modulation $\Delta \phi_k > \pi$ which affects the complete two-dimensional phase-plane $(\theta, \psi)$. 
While the first method depends on the folded shape of the stable manifold of $S$ as discussed in Section \ref{Sect::Manifolds}, the second one does not and has a operating region which can be ten times as large.\\
Method $1$ is shown in Fig.\ \ref{Fig::PhaseMod}(a)-(c). Starting from the clockwise ($CW$) OPAS solution for $\phi_{k, i} = 0.7$ ($OPAS_{\phi i, CW}$), we change the linear coupling phase to $\phi_{k, f} = 1.5$, such that $OPAS_{\phi i, CW}$ now finds itself in the basin of attraction of the counter-clockwise ($CCW$) OPAS state ($OPAS_{\phi f, CCW}$) for the dynamical phase picture with $\phi_{k, f} = 1.5$. In this new situation, the system follows first the stable and then the unstable manifold of the saddle point $S_{\phi f}$, such that it finally relaxes to the stable $OPAS_{\phi f, CCW}$ state. If we then remove the modulation of the coupling phase, the system feels again the manifolds of the original phase-space picture ($\phi_{k, i} = 0.7$). Since the $OPAS_{\phi f, CCW}$ state finds itself in the basin of attraction of the $CCW$ OPAS state for $\phi_{k, i} = 0.7$, it relaxes to this $OPAS_{\phi i, CCW}$ state, such that we have accomplished a successful switch. \\
Since we have seen in Sect. \ref{Sect::Manifolds} that the spiraling shape of the stable manifold is quite sensitive to parameter variations, the necessity of this folding shape for the directional switching to occur can be a drawback of this method. Therefore, we propose a second scheme, which is less demanding on the particular phase space structure of the reduced dynamical system.
In Fig.\ \ref{Fig::PhaseMod}(b)-(d), we again prepare the system in the $OPAS_{\phi i, CW}$ state with $\phi_{k, i} = 0.7$, but now we change the coupling phase to $\phi_{k, f} = 4.3$. In doing this, the $OPAS_{\phi i, CW}$ state is now located in the basin of attraction of the in-phase asymmetric solution $IPAS_{\phi f, CW}$, such that it relaxes to that state. Furthermore, this $IPAS_{\phi f, CW}$ state finds itself in the basin of attraction of the counter-propagating mode $OPAS_{\phi i, CCW}$. By removing the modulation ($\phi_{k, i} = 0.7$), the system thus follows first the stable and then the unstable manifold of the saddle $S_{\phi i}$ to end up having successfully switched to the CCW mode $OPAS_{\phi i, CCW}$. It is important to note that this second method works for $\phi_{k, f} \in [4.27 - 4.47]$ (for this particular  $\phi_{k, f}$ and current $J$), which is a more than ten times bigger operating region than for method 1.

\section{Discussion and Concluding remarks}\label{Sec::Discussion}

To summarise, in this paper we have presented a complete stability analysis of a SRL which is based on an asymptotic two-dimensional reduction of a rate-equation model \cite{VanderSandeJPhysB2008}. 
In particular, we have found the global bifurcation curves corresponding to a homoclinic bifurcation of limit cycles and fold of cycles [See Fig.~\ref{fig:StabilityDiagram}] which are responsible for the global shape of the invariant manifolds of the system. Four different Takens-Bogdanov codimension-2 points have been located, around which the dynamics of the system is organized.

For parameters which correspond to directional bistability, the shape of the basins of attraction has been investigated. The basin of attraction of the CW and CCW are separated by the stable manifold of the saddle $S$, and fold one into each other. The role of the different physical parameters of the system in the folding/unfolding of the manifold has been addressed.

After our topological analysis, we have proposed two novel schemes to switch the operation of the ring from clockwise (CW) to counter-clockwise (CCW). Such schemes relying on the folding shape of the invariant manifolds of the system, are alternative to the standard approach of injecting a counter-propagating signal \cite{YuanIEEE2008, PerezOE2007, Gelens_OL_2008}.

The first of such schemes consists of a modulation in the bias current, which drags the system outside the basin of attraction of the initial state and releases it when inside the basin of attraction of the final state. 
Our simulations reveal the existence of a limiting fall time $t_{fall}$ for the applicability of the scheme. Although the switching speed, obtainable with this scheme,  is of course limited by the electrical bandwidth of the modulation source, all-optical methods relying on phase space steering \cite{Lippi2000} can be applied to increase the performance.

A second scheme has been proposed which consists of modulating the backscattering phase $\phi_k$. We discuss switching takes place in two different ways when the amplitude of the phase modulation is larger or smaller than $\pi$. Numerical simulations have demonstrated the applicability of the scheme in both cases.

As a final remark, we want to state that this phase-space topology is common for optical systems, for instance a nonlinear ring resonator described by the Ikeda map \cite{Hammel85a}. Moreover, our reduced two-dimensional SRL model can --- in the region close to the bidirectional solution ($\theta \approx 0$) --- also be seen as an optical prototype of the symmetric FitzHugh-Nagumo equation, which describes excitable systems, and is important in the study of neurons \cite{Fitzhugh_1961, Izhikevich_2000}. More generally, we expect our two-dimensional system to be relevant for planar dynamical systems which are $Z_2$-invariant \cite{Kuznetsovbook}, such that the techniques discussed here could be applicable to more and diverse dynamical systems.

\begin{acknowledgments}
This work has been funded by the European Community under project IST-2005-34743 (IOLOS) and the Belgian Science Policy Office under grant No.\ IAP-VI10. LG  is a PhD Fellow, while G.V. and S.B. are Postdoctoral Fellows of the Research Foundation - Flanders (FWO). Also project support from FWO is acknowledged. Furthermore, we thank Pere Colet, Alessandro Scir\'e, Manuel Mat\'ias, Marc Sorel, Guido Giuliani, Siyuan Yu and Kirk Green for interesting discussions.
\end{acknowledgments}


\end{document}